# Towards fast quantum key distribution with quantum frames

I. Lucio Martinez (1), P. Chan (1), X. Mo (1), S. Hosier (2), and W. Tittel (1)*

1: University of Calgary, 2500 University Drive NW, Calgary, Alberta, Canada T2N 1N4
2: SAIT Polytechnic, 1301 16th Ave NW, Calgary, Alberta, Canada T2M 0L4

**Abstract**

We propose and investigate a fibre-based quantum key distribution system, which employs polarization qubits encoded into faint laser pulses. As a novel feature, it allows sending of classical framing information via sequences of strong laser pulses that precede the quantum data. This allows synchronization, sender and receiver identification, and compensation of time-varying birefringence in the communication channel. Furthermore, this method also provides a platform to communicate implementation specific information such as encoding and protocol in view of future optical quantum networks. All optical components can be operated at Gbps rates, which is a first requirement for Mbps secret key rates.

## Introduction

Based on the particular properties of single quantum systems, quantum key distribution (QKD) promises cryptographic key exchange over an untrusted, authenticated public communication channel with information theoretic security [1]. While commercial QKD systems have existed for a few years [2], their employment is currently limited by small key rates and reduced distance, and the integration into networks is still at an early stage. In this paper, we present the current status of our QKD system, which aims at delivering secret keys at Mbps rates in a real-world fibre network environment over a distance of around ten kilometres.

## The QKD system

Our proposed QKD system is based on polarization qubits, and employs the BB84 protocol [3], supplemented with two decoy states [4]. It allows alternating sequences of strong and faint laser pulses, encoding classical data and quantum data, respectively (see Fig. 1). The classical data forms the *control frame (C-frame)*, and a pair of classical/quantum data forms a *quantum frame* (*Q-frame*). The C-frame allows synchronizing sender Alice and receiver Bob, facilitates time-tagging, and provides a platform to communicate sender and receiver address (for routing or packet switching) plus implementation specific information such as encoding (e.g. polarization or time-bin qubit [5]) and protocol (e.g. BB84 [3], decoy state [4], or B92 [6]). This is interesting in view of reconfigurable networks comprising different QKD technologies. The classical information is encoded into specific polarization states, allowing assessment and compensation of time-varying birefringence in the quantum channel. Note that the compensation scheme can easily be adapted to other QKD setups employing e.g. time-bin qubits, entanglement, or quantum repeaters.

A schematic of the QKD system is depicted in Fig. 2. At Alice's, two laser diodes (1550 nm wavelength, max. trigger rate > 1GHz, from Avanex) with polarization maintaining fibre pigtails are used to generate the C-frame ($LD_C$) and the quantum data ($LD_Q$). The pulses emitted from $LD_Q$ are first attenuated by a fixed attenuator (ATT), and then sent through a 10 Gbps intensity modulator (IM, from Avanex) to create signal and decoy states with mean photon number per pulse $\mu$ of 0.5 and 0.1, resp. To create vacuum decoy states ($\mu=0$), no electrical pulses are sent to $LD_Q$. The horizontally polarized faint pulses are then transmitted through PBS1 (from OZ-Optics), and combined with the strong, vertically polarized pulses from $LD_C$. All pulses are then sent to a polarization modulator, a 10 Gbps LiNbO$_3$ phase modulator whose active axis is aligned at 45° with respect to the input polarization states (from EOspace). Depending on the applied voltage, horizontal (H), vertical (V), right (R), or left (L) circular polarization states can be created.

After being transmitted through a fibre optics quantum channel to Bob, 10% of the light will in future be directed towards a fast photo detector (DET) followed by a logic device (LOG), which reads the information encoded into the C-frame and takes appropriate action [7]. The remaining light is split at a 50/50 beam splitter (BS, from Thorlabs), and directed to two polarization stabilizers (PS1, PS2, from General Photonics) followed by polarization beam splitters (PBS2, PBS3, from General Photonics) and single photon detectors (SPDs, from idQuantique). PS1 ensures that horizontally polarized C-frames, and hence qubits, emitted at Alice's arrive unchanged at PBS2. Similarly, PS2 is set up such that right circular polarized C-frames and qubits emitted at Alice's always impinge horizontally polarized on PBS3. Since the transformation in the communication channel is described by a unitary matrix (i.e. orthogonal states remain orthogonal), our stabilization scheme ensures that qubits prepared in H and V, or R and L states arrive horizontally and vertically polarized on PBS2 or PBS3, respectively. Hence, the two sets of PS, PBS and SPDs (a) allow compensation of unwanted polarization transformations in the quantum channel [8], and (b) allow projection measurements onto H, V, R and L, as required in the BB84 protocol.

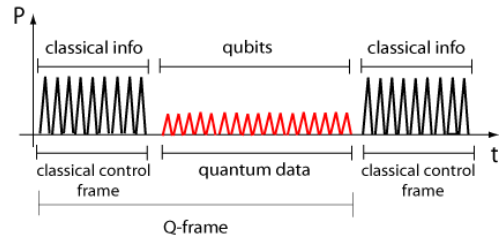

*Figure 1: Quantum framing with alternating classical control frames (inspired by the Ethernet protocol) and quantum data. The schematic shows optical power P versus time t.*

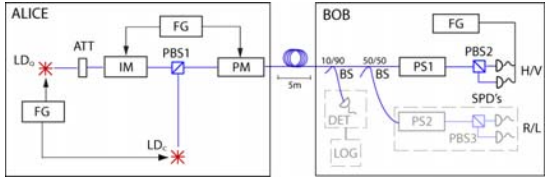

Figure 2: Schematic of our QKD system. The gray, dashed drawn components have not been implemented in the proof-of-principle demonstration. $LD_Q$ and $LD_C$ - laser diodes; ATT - attenuator; IM - intensity modulator; PM - polarization modulator; BS - beam splitter; PS - polarization stabilizer; PBS - polarization beam splitter; D – photo detector; LOG – logic device; SPD - single photon detector; FG - function generator.

**The proof-of-principle demonstration**

We tested our QKD setup as follows. First, we automatically compensated the polarization transformation in a 5 m long fibre connecting Alice and Bob with the help of a 10 sec long pulse from $LD_C$ (representing a simple C-frame) and PS1 [9]. We then generated horizontally polarized faint pulses with the three aforementioned mean values of photons per pulse and measured the quantum bit error rate (QBER), see table 1. We continued the experiment for vertically polarized qubits, and, after repeating the compensation step for the second set of states, with right and left polarized qubits. The average QBER for the signal states with $\mu=0.5$ is 3.4%, of which we attribute 3% to imperfect qubit generation, 0.2% to imperfect compensation of birefringence in the quantum channel, and 0.2% to detector noise.

**Conclusion and outlook**

We have proposed a novel, fibre-based QKD system employing polarization encoding and quantum frames, and have demonstrated a proof-of-principle, showing that polarization information encoded in the classical control frames can indeed be used to stabilize unwanted qubit transformation in the quantum channel. All optical elements in our setup can be operated at Gbps rates, which is a first requirement for a future system delivering secret keys at Mbps. In order to remove another bottleneck towards a high rate system, we are investigating forward error correction based on Low Density Parity Check Codes [10]. Work on a field implementation of the complete system over a 12 km dark fibre between the University of Calgary and SAIT Polytechnic is under way.

Table 1: QBER for different qubit states and mean photon numbers $\mu$. Statistical errors are 0.01% for $\mu=0.5$, and 0.04% for $\mu=0.1$. The QBER for the vacuum state $\mu=0$ is $(0.15\pm 0.004)$%.

| qubit | H | | V | | R | | L | |
|---|---|---|---|---|---|---|---|---|
| photon number | 0.5 | 0.1 | 0.5 | 0.1 | 0.5 | 0.1 | 0.5 | 0.1 |
| QBER | 4.1 | 6.2 | 3.3 | 4.8 | 4.2 | 5.6 | 2.2 | 5.4 |

This work is supported by General Dynamics Canada, iCORE, NSERC, QuantumWorks, CFI, AET, CMC Microsystems, and CONACYT.